\documentclass[journal=ancac3,manuscript=article,layout=twocolumn
]{achemso}

\usepackage[version=3]{mhchem} 

\author{Francis Granger}
\email{francis.granger@cea.fr}
\affiliation{Univ. Grenoble-Alpes, CEA, Grenoble INP, IRIG, PHELIQS, NPSC, 38000 Grenoble, France.}
\affiliation{Univ. Grenoble-Alpes, CNRS, Inst. NEEL, 38042 Grenoble, France.}

\author{Edith Bellet-Amalric}
\affiliation{Univ. Grenoble-Alpes, CEA, Grenoble INP, IRIG, PHELIQS, NPSC, 38000 Grenoble, France.}

\author{Kuntheak Kheng}
\affiliation{Univ. Grenoble-Alpes, CEA, Grenoble INP, IRIG, PHELIQS, NPSC, 38000 Grenoble, France.}

\author{Gilles Nogues}
\author{David Ferrand}
\author{Joël Cibert}
\email{joel.cibert@neel.cnrs.fr}
\affiliation{Univ. Grenoble-Alpes, CNRS, Inst. NEEL, 38042 Grenoble, France.}

\title{Calibration-free measurement of the phonon temperature around a single emitter}

\keywords{phonon coupling, temperature, quantum emitter, thermometer, laser heating, zero-phonon line, phonon sideband}

\begin{document}


\begin{abstract}
The emission properties of a localized solid-state emitter are strongly influenced by its environment. The coupling to acoustic phonons impacts the coherence of the emitter and its temperature dependence, and also results in the apparition of phonon sidebands besides the sharp zero-phonon line. Here, we present a method for measuring the absolute temperature of a localized emitter by directly plotting the ratio of the Stokes and anti-Stokes components of the phonon sideband as a function of the shift from the zero-phonon line. This approach requires no calibration and knowledge of the system, making it applicable to a wide range of emitters and materials. We validate the method using a CdSe quantum dot in a ZnSe nanowire. We thus show that the quantum dot is significantly heated under non-resonant excitation when increasing the incident power at low temperature and is ascribed to the drop in thermal conductivity at these temperatures.
\end{abstract}


\vspace{2cm}

Solid-state quantum emitters are actively studied for applications in a wide range of quantum technologies, including quantum communication, sensing, and computing. However, these sources are inevitably coupled to their immediate environment, significantly affecting their performance. For instance, the interaction with phonons introduces decoherence, reducing the indistinguishability of the single photons \cite{denning_phonon_2020,grange_reducing_2017,reigue_probing_2017}. This leads to a trade-off between brightness and purity at elevated temperatures \cite{granger_brightness_2023}. 

\begin{figure*}
  \centering
   \includegraphics[width=\textwidth]{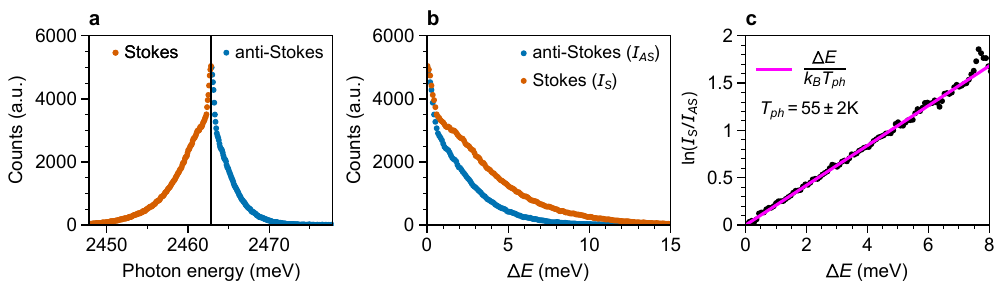}
  \caption{
(a) Photoluminescence intensity spectrum at $T_\mathrm{cryo}=50\,\mathrm{K}$ under $P=14\,\mu\mathrm{W}$ continuous wave excitation power. The Stokes intensity $I_{AS}$ (orange dots) and anti-Stokes intensity $I_{AS}$ (blue dots) are separated by the central position of the ZPL, shown by a vertical black line. (b) Both sidebands are plotted as a function of the distance $\Delta E$ to the ZPL. (c) The computed logarithm of the ratio ($I_S/I_{AS}$) is compared to the expected expression $\Delta E/(k_\mathrm{B}T_{\mathrm{ph}})$.
}
  \label{fig:fig1}
\end{figure*}

The necessity to understand and control the influence of phonons has increased the need for high-accuracy temperature measurements at cryogenic temperatures. The temperature in the vicinity of the quantum emitter is usually assumed to be close to that of a reference thermometer attached to the sample holder. 
However, temperature can vary significantly across a material, and traditional, bulk measurements fail to capture local temperature gradients at a nanometer scale. 

The objective of this paper is to demonstrate a method to accurately measure a local temperature, without the need of any calibration. Furthermore, the technique is particularly interesting in the context of single photon emission, since it measures specifically the temperature of the phonons which are actually interacting with the quantum emitter.

\begin{figure*}
  \centering
   \includegraphics[width=\textwidth]{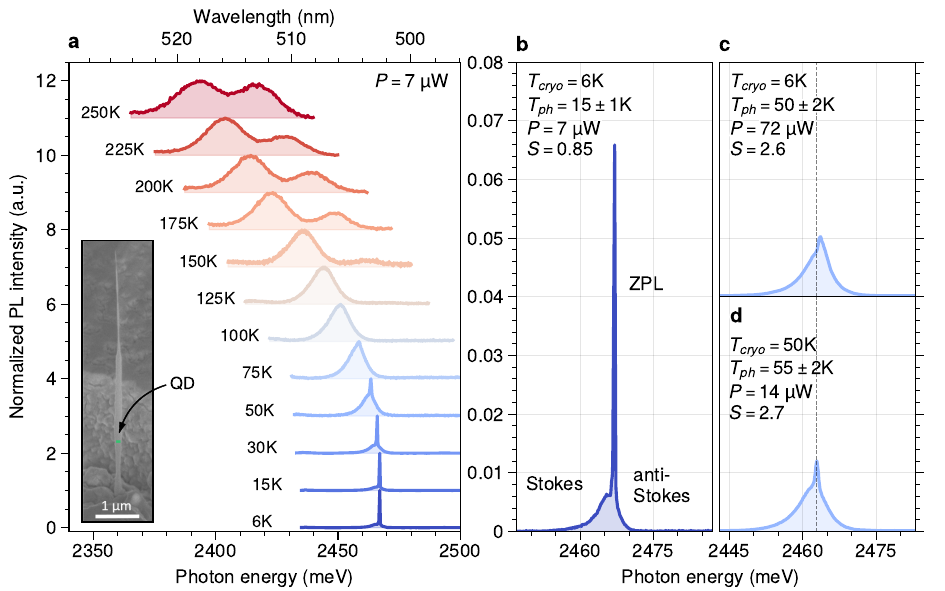}
    \caption{
(a) PL spectra (normalized in intensity) recorded at different cryostat temperatures under continuous wave excitation (\(P=7\,\mu\mathrm{W}\)) at \(405\,\mathrm{nm}\). An offset between spectra is added for clarity. The inset shows the QD-NW emitter discussed in this article, with the green rectangle marking the expected position of the quantum dot. (b, c) PL spectra (all normalized to the same integrated intensity) for various excitation powers and cryostat temperatures.
}
  \label{fig:fig2}
\end{figure*}

The method exploits the phonon sidebands which appear in the photoluminescence (PL) spectrum or a solid-state emitter embedded within a crystalline host due to the interaction with phonons. In particular, the interaction with acoustic phonons adds a broad phonon sideband on both sides of the sharp zero-phonon line (ZPL) \cite{besombes_acoustic_2001,krummheuer_theory_2002}, as observed in Fig.~\ref{fig:fig1} and \ref{fig:fig2} for a CdSe quantum dot (QD). The lower-energy (phonon emission) and higher-energy (phonon absorption) regions of the phonon sideband are referred to as the Stokes and anti-Stokes sidebands, respectively. The phonon sideband  lineshape and its intensity strongly depend on the temperature.

Besombes \textit{et al.} \cite{besombes_acoustic_2001}  first observed and explained this effect in CdTe QDs by extending the Huang-Rhys theory of localized electron-phonon interaction in color centers \cite{huang_theory_1997} to the exciton system in a QD. Further studies revealed similar lineshapes in III-V QDs based on InAs \cite{favero_acoustic_2003, reigue_boite_2017}, GaAs \cite{peter_phonon_2004}, or GaN \cite{rol_probing_2007}. More recently, phonon sidebands have also been studied in emerging two-dimensional materials, including in WSe\textsubscript{2} \cite{vannucci_single-photon_2024} and hexagonal boron nitride (hBN) \cite{khatri_phonon_2019} single photon emitters.

Our method is based on the extraction of temperature from the population of the acoustic phonons that contributes to the complex lineshape of a localized optical emitter. While the method is general and can be applied to diverse emitters and materials, we validate it here with experimental data obtained from PL measurements on a single CdSe QD embedded in a ZnSe nanowire \cite{gosain_room_2021}. Using this system, as summarized in Fig.~\ref{fig:fig1}, we demonstrate the ability to measure the local temperature of acoustic phonons relevant for the emitter properties. We also explore the heating effects induced by non-resonant optical excitation. The analysis provides a quantitative overview of the relationship between excitation power, thermal conductivity, and local phonon temperature. More information concerning the CdSe-ZnSe system and the experimental setup used to measure the PL spectra are provided in the Methods section \ref{sec:methods}.

\section{Absolute temperature from the acoustic-phonon sidebands}
\label{sec:phonon_temp}

The absorption and emission of a phonon, labelled by $i$, are governed by its occupation $n_i$. These processes are driven from phonon-assisted luminescence. The probability of a Stokes process involving the emission of a single-phonon is proportional to $(n_i+1)$, while that of an anti-Stokes process involving the absorption of a single phonon is proportional to $n_i$. Details of the mechanisms are contained in the prefactor, which is the same in both cases, so that the ratio of the Stokes to anti-Stokes intensities is $(n_i+1)/n_i$. 

If thermal equilibrium is achieved, the population is given by the Bose-Einstein distribution:

\begin{equation}
    n_i=n(\hbar\omega_i,T_{\mathrm{ph}}) = \frac{1}{\exp \left(\frac{\hbar\omega_i}{k_\mathrm{B}T_{\mathrm{ph}}} \right) - 1},
\end{equation}

\noindent
where $\hbar\omega_i$ is the phonon energy, $k_\mathrm{B}$ the Boltzmann constant, and $T_{\mathrm{ph}}$ the phonon temperature. Then the Stokes to anti-Stokes ratio for a single phonon process is $\exp(\frac{\hbar \omega_i}{k_BT_{ph}})$. 

At low temperatures (\(T_{\mathrm{ph}} \ll \hbar\omega_i/k_\mathrm{B}\)), the phonon populations are small. Therefore, the dominant process is the emission of a single phonon which results in a non-vanishing Stokes band, and a strong asymmetry with the almost-vanishing anti-Stokes sideband.



At higher temperature, multi-phonon contributions appear. A contribution involving a set of phonons $\{\hbar\omega_i\}$ emitted, and $\{\hbar\omega_j\}$ absorbed, with $\Delta E=\Sigma_i \hbar\omega_i-\Sigma_j \hbar\omega_j>0$, appears at $\Delta E$ below the ZPL (Stokes process). The opposite phonon path with $\{\hbar\omega_i\}$ phonons absorbed and $\{\hbar\omega_j\}$ emitted, contributes to the anti-Stokes band at the same $\Delta E$ above the ZPL.

For such a set of phonons, the ratio of the contributions to the Stokes and anti-Stokes sidebands can be written as

\begin{multline}
\label{eq:ratio}
\prod_i \frac{n(\hbar\omega_i)+1}{n(\hbar\omega_i)}
\prod_j \frac{n(\hbar\omega_j)}{n(\hbar\omega_j)+1}
    = \frac{\prod_i \exp\left(\frac{\hbar\omega_i}{k_B T_{\mathrm{ph}}}\right)}{\prod_j \exp\left(\frac{\hbar\omega_j}{k_B T_{\mathrm{ph}}}\right)}\\
= \exp\left(\frac{\Delta E}{k_B T_{\mathrm{ph}}}\right).
\end{multline}

This is valid for each contribution at $\Delta E$, hence for the overall intensity ratio of the Stokes and anti-Stokes bands \(I_S (E_{ZPL}-\Delta E)/I_{AS}(E_{ZPL}+\Delta E)\). More details on the derivation of Eq.~\ref{eq:ratio} can be found in the Methods section\ref{sec:methods}.

The key to the method lies in plotting this Stokes to anti-Stokes intensity ratio, \(I_S (E_{ZPL}-\Delta E)/I_{AS}(E_{ZPL}+\Delta E)\), for all values of the shift $\Delta E$ to the zero-phonon line. 

This approach provides a direct and absolute measure of the local temperature near the emitter, regardless of the number of phonons involved in the optical transitions. The temperature value obtained is independent of the parameters which govern the lineshape, such as the coupling strength, the shape and size of the exciton envelope, and the phonon density of states. Therefore, the method is versatile and suitable for many systems and materials. If the line is well isolated from spurious contributions, the method can be applied at elevated temperatures as long as the sideband asymmetry is clearly distinguishable, and at low temperature as long as the sideband signal significantly emerges from the background noise. The data analysis method is described in more detail in the Methods section.

\section{Results and discussion}
\label{sec:results}

The method is demonstrated and validated for two important systems: single QDs and a 2D quantum emitter. Specifically, we examine a CdSe QD embedded in a ZnSe nanowire \cite{gosain_room_2021,granger_brightness_2023} and an InAs/GaAs QD \cite{reigue_boite_2017}, as well as a $\text{WSe}_2$ quantum emitter \cite{vannucci_single-photon_2024}. This section focuses on a single CdSe QD within a ZnSe nanowire. Complementary results for the InAs/GaAs QD and the $\text{WSe}_2$ quantum emitter are provided in the Supporting Information.

\subsection{Application to a CdSe QD in a ZnSe nanowire}

The PL spectrum of a CdSe QD measured at the cryostat temperature $T_{\mathrm{cryo}} = 50\,\mathrm{K}$ under a non-resonant excitation power of $P = 14\,\mu\mathrm{W}$ is shown in Fig.~\ref{fig:fig1}a. Details on the experimental setup are presented in section Methods \ref{sec:methods}. 

At this temperature, the phonon sideband is asymmetric and the ZPL can be easily distinguished from the Stokes/anti-Stokes sidebands. The intensities of both sidebands ($I_S$ and $I_{AS}$) are plotted with respect to the ZPL position $E_{ZPL}$ in Fig.~\ref{fig:fig1}b. The logarithm of the Stokes to anti-Stokes ratio ($I_S/I_{AS}$) is displayed in Fig.~\ref{fig:fig1}c. This analysis covers energy shifts $\Delta E$ up to $8\,\mathrm{meV}$. 

The straight line observed in Fig.~\ref{fig:fig1}c indicates the existence of a single and uniform temperature across all phonon modes coupled to the system. This result demonstrates the thermal equilibrium of acoustic phonons within this energy range. From the slope, we determine the unique phonon temperature to be $T_{\mathrm{ph}} = 55 \pm 2\,\mathrm{K}$, which is significantly higher than the cryostat temperature.

The same method was applied to spectra taken at various cryostat temperatures, which are shown in Fig.~\ref{fig:fig2}a. At temperatures up to $100$~K, the spectrum features a single emission line (ZPL+phonon sideband). However, as the temperature increases ($T_{\text{cryo}}>125$~K), an additional contribution emerges at higher energy. From prior measurements on the same sample \cite{granger_brightness_2023}, including polarized spectra, the single contribution observed at cryogenic temperatures is attributed to the charged exciton, and the contribution at higher energy to the neutral exciton. The presence of this additional contribution sets the upper limit for applying the method to higher temperatures in the present sample. 

Figure~\ref{fig:fig2}b shows the charged exciton line emission spectrum taken at $T_\mathrm{cryo}=6\,\mathrm{K}$, which is the lowest temperature achievable by our cryostat. At this temperature, the phonon sideband is strongly asymmetric with respect to the ZPL position $E_{ZPL}$. The ZPL contributes for 43\% of the whole spectrum. The Huang and Rhys factor $S$ (which is a measure of the average number of phonons involved in a transition) is obtained by computing the ratio of the ZPL intensity to the whole line: $\exp{(-S)}=I_{ZPL}/I_\text{tot}$. Therefore, at $T_\mathrm{cryo}=6\,\mathrm{K}$,  $S\approx0.85$. From the Stokes to anti-Stokes ratio, we measure a temperature $T_{\mathrm{ph}} = 15 \pm 1\,\mathrm{K}$, much higher than the temperature of the cryostat.

When increasing the temperature, the intensity of the ZPL diminishes drastically in favor of the sideband as phonon interactions redistribute the emission energy into the phonon replica. At $T_{\text{cryo}}=50\,\mathrm{K}$ ($T_{\mathrm{ph}} = 55 \pm 2\,\mathrm{K}$ as shown in Fig.~\ref{fig:fig1}c and \ref{fig:fig2}d), the ZPL accounts for only 7\% of the total intensity, \textit{i.e.}, $S=2.7$: More multi-phonon processes are involved. Figure~\ref{fig:fig3}b displays the dependence of the phonon temperature on cryostat temperature (with excitation power held constant). At higher cryostat temperatures ($T_{\text{cryo}} > 50 \, \text{K} $), the temperature measured in the vicinity of the QD approaches the cryostat’s temperature, with $ T_{\text{ph}} \approx T_{\text{cryo}} $, indicating that thermal equilibrium occurs more efficiently at these temperatures.

Above a temperature of $T_{\text{cryo}}=100\,\mathrm{K}$, the ZPL can no longer be resolved from sidebands, and the phonon sideband becomes nearly symmetric. Yet, a local phonon temperature of $T_\mathrm{ph}=98\pm5\,\mathrm{K}$ can be measured.

A similar evolution of the spectral lineshape is observed when increasing the excitation power at constant cryostat temperature. Figure~\ref{fig:fig2}c shows the spectrum recorded at low cryostat temperature, $T_\mathrm{cryo}=6\,\mathrm{K}$, but under a high excitation power $P=72\,\mu\mathrm{W}$. The local temperature was found to be $T_\mathrm{ph}=50\pm2\,\mathrm{K}$, much higher than the cryostat temperature and corroborating the visual similarities between the two sidebands observed in Fig.~\ref{fig:fig2}c and \ref{fig:fig2}d. This is further confirmed with the Huang and Rhys $S$ factor, which is equal to that obtained at $T_\mathrm{cryo}=50\,\mathrm{K}$ and low power: Actually, the temperature dependence of $S$ can be another method to measure the temperature, but it requires calibration \cite{rudin_temperature_2006}. 

This points to a significant laser heating when the emitter is excited at a higher power level. This is confirmed in Fig.~\ref{fig:fig3}c which shows that the phonon temperature measured at $T_\mathrm{cryo}=6\,\mathrm{K}$ strongly depends on the excitation power.

\subsection{Laser heating of a quantum dot in a nanowire}

The effect of localized laser heating has been evidenced on various color centers, using the redshift of the ZPL, and modeled \cite{misiara_effects_2024,dharmasiri_sensitivity_2024,gao_local_2024}. This section derives a simple model to account for the heating effect observed in our case study.

\begin{figure*}
  \centering
   \includegraphics[width=\textwidth]{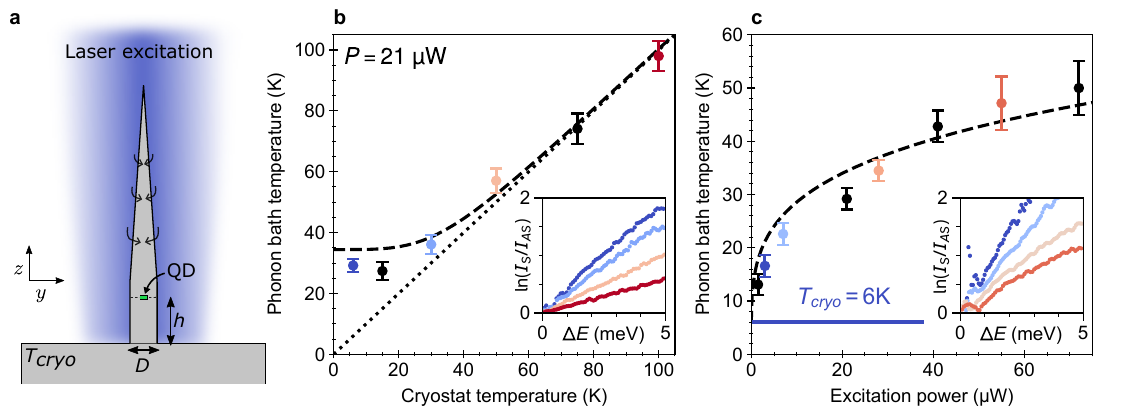}
    \caption{
(a) Schematic of the laser beam heating of the QD-NW. 
(b) Measured phonon temperature as a function of the cryostat temperature. The black dashed line represents the temperature at the QD calculated with the profile model for an excitation power $ P = 21 \, \mu\text{W} $, while the black dotted line corresponds to $ T_{\text{ph}} = T_{\text{cryo}} $. 
(c) Measured phonon temperature at various excitation powers for $ T_{\text{cryo}} = 6 \, \text{K} $, with the result of the model shown by the dashed line. The logarithm of the Stokes to anti-Stokes ratios for a few temperature values (identified by the color of symbols) are displayed in the inset of (b) and (c).
}
  \label{fig:fig3}
\end{figure*}

The emitter investigated within this work was grown by molecular beam epitaxy on a (111)B GaAs substrate with a nm-thick ZnSe buffer layer. It consists of a vertical ZnSe nanowire embedding a CdSe quantum dot near its tip \cite{raj_gosain_onset_2022,gosain_room_2021}, encapsulated within a thick ZnSe shell. The shell acts as a photonic waveguide and enhances the collection efficiency \cite{claudon_highly_2010}. Alongside the nanowire growth, a two-dimensional ZnSe layer was also formed on the substrate. The nanowires typically show lengths of 5–6 µm and a maximum diameter of approximately 200 nm, while the 2D layer has a thickness of around 1 $\mu\text{m}$. A schematic representation of the system is shown in Fig.~\ref{fig:fig3}a and a SEM image the QD-NW system is shown in Fig~\ref{fig:fig2}a inset.

\subsubsection{Power balance}

The microscope focuses the laser beam through a glass window onto the vertical nanowire placed in vacuum with a pressure below $10^{-6}$ mbar. 

The nanowire receives a constant power density $ \frac{P}{S_{\text{beam}}} $, where $P$ is the power of the incident beam and $S_{\text{beam}}$ is its area. The absorption takes place over the whole nanowire’s length. According to Zhan et. al \cite{zhan_enhanced_2014}, it absorbs a power  $P\frac{S_{NW}}{S_{\text{beam}}}{Q_{\text{abs}}}$ with $S_{NW}$ the cross-sectional area of the nanowire at the level of the QD, and $Q_{\text{abs}}$ is an absorption coefficient that accounts for the geometry of the system. For a tapered photonic wire, $Q_{\text{abs}} \sim 10$. Our beam area is $S_{\text{beam}} = 1.5 \, \mu\text{m}^2$,  and $S_{NW}= 0.03 \, \mu\text{m}^2$ from the SEM image shown in the inset of Fig~\ref{fig:fig2}a, hence about 20\% of the incident power is absorbed by the NW.

The vacuum environment eliminates air conduction and thermal convection losses. The energy losses due to the ZnSe luminescence are expected to be low and the thermal radiation losses are negligible (see Methods).


Therefore, under these assumptions, the absorbed power is mainly converted into phonons, dissipating through thermal conduction to the substrate and the sample holder.

\subsubsection{Temperature gradient}

The QD is located near the base of the NW at a distance $h$, and the NW around it and below can be modelled as a cylinder with cross-section $S_{NW} = \frac{\pi D^2}{4}$ heated from the top (as shown schematically in Fig. \ref{fig:fig3}a). The typical phonon dynamics are expected to occur on the microsecond timescale, allowing the system to be treated in a steady-state regime. In this regime, the heat flux remains constant over time, and the temperature can be considered uniform across the diameter of the nanowire. Thus, we get from the one-dimensional heat transport equation along the nanowire axis $z$:

\begin{equation}
    \kappa(T) S_{NW} \frac{dT}{dz} = P\frac{S_{NW}}{S_{\text{beam}}} Q_{\text{abs}},
\end{equation}

where $\kappa(T)$ is the thermal conductivity at temperature $T$. At low temperature, the thermal conductivity takes the form $\kappa=1/3Cv_s\Lambda$ with $C$ the volumetric heat capacity of the material, $v_s$ the phonon velocity (mainly acoustical phonons) and $\Lambda$ their mean free path. The phonons exhibit ballistic transport behavior, with a phonon mean free $\Lambda$ path limited by the characteristic dimensions of the wire \cite{ziman_electrons_2001,bourgeois_measurement_2007}. The thermal conductivity is reduced with respect to the bulk value by phonon boundary scattering. Within this so-called Casimir regime, the thermal conductivity follows a cubic power law $\kappa = \alpha T^3$.

In the absence of data for ZnSe nanowires, we adapted the experimental data from silicon nanowires from Bourgeois \textit{et al.} \cite{bourgeois_measurement_2007} to estimate the low-temperature thermal conductivity of a nanowire with a diameter of $200 \, \text{nm}$. For the nanowire system studied, we use an approximate coefficient $\alpha \approx 4 \times 10^{-4} \, \text{W m}^{-1} \text{K}^{-4}$.

Solving the heat equation with $\kappa = \alpha T^3$ leads to a straightforward expression for the temperature profile along the nanowire. For a QD positioned at a height $h$ above the surface, and assuming that the substrate is at $T_{cryo}$, the temperature at $h$ is given by:

\begin{equation}
T(h) = T_{cryo} \left( 1+\frac{4 P Q_{\text{abs}}}{\kappa (T_{cryo}) T_{cryo} S_{\text{beam}}} h \right)^{\frac{1}{4}}.
\end{equation}

The beam area is taken to be $S_{\text{beam}} = 1.5 \, \mu\text{m}^2$,  and from the SEM image shown in the inset of Fig~\ref{fig:fig2}a, the QD is positioned at a height $h = 1 \, \mu\text{m}$ above the surface. These parameters were used to calculate the dashed lines in Fig.~\ref{fig:fig3}b and Fig.~\ref{fig:fig3}c.

It is worth noting that some of the parameters used in this simplified model, such as the thermal conductivity and absorbed power, involve significant uncertainties. For instance, the thermal conductivity is highly sensitive to the nanowire’s geometry, dimensions, and surface characteristics, which are difficult to quantify precisely. Similarly, the absorbed power at the position of the QD is influenced by factors like the nanowire’s geometry, optical properties, and excitation wavelength. Despite this, this simple model satisfactorily reproduces the trends observed.

The small value of the thermal conductivity $\kappa (T_{cryo})$ is responsible for the strong initial dependance on excitation power at low temperature $T_{\text{cryo}} = 6 \, \mathrm{K}$ in Fig.~\ref{fig:fig3}c, and the difference of the temperature at the QD with respect to the cryostat temperature in Fig.~\ref{fig:fig3}c. As the overall temperature increases, the thermal conductivity rapidly increases and the temperature difference drops to zero in Fig.~\ref{fig:fig3}b. The increase in thermal conductivity also explains the smoother dependance at high power in Fig.~\ref{fig:fig3}c. It similarly affects the dependance on $h$ so that the temperature calculated at the tip of the NW is not significantly higher than the temperature at the level of the QD.


\subsection{A universal phonon thermometer}

The method presented here is reminiscent of the well-established Raman spectroscopy technique, which allows a measure of the optical phonon temperature of materials by analyzing the Stokes and anti-Stokes components of the Raman spectrum \cite{schrader_infrared_2008,moore_raman_2014}. However, in our case, we probe the thermal properties of phonons interacting directly with the emitter. In particular, these phonons will impact the ZPL characteristics and the use of the emitter in quantum technology \cite{denning_phonon_2020,grange_reducing_2017}. Other methods such as spin polarization studies \cite{zeman_boltzmann_2024} allow the measurement of local temperatures in crystals, but the technique requires a calibration.

The uncertainty in the temperature extraction mainly arises from the background signal and the choice of the zero-phonon line position ($E_{ZPL}$), as slight variations in the position can significantly affect the calculated temperature. Although the ZPL is included in the computation, for sufficiently narrow linewidths, the ZPL has minimal impact on the Stokes-to-anti-Stokes ratio. Additionally, it can be noted that the spectrum taken at higher excitation power in Fig.\ref{fig:fig2}d presents a broader ZPL, which we ascribe to spectral diffusion due to increased charge fluctuations of photo-created carriers surrounding the QD. At constant temperature, the ratio plot is consistently more noisy at higher power excitation (see insert of Fig.~\ref{fig:fig3}c. This might be due to the broader ZPL linewidth, but also to additional, parasitic lines superimposed onto the sidebands.

The CdSe QD studied throughout this paper was particularly suited for studying phonon interactions due to a unique and bright spectral line that remains stable and well isolated over a wide temperature range, from 6 K to 100 K, as shown in Fig.~\ref{fig:fig2}. Single QDs in nanowires \cite{gosain_room_2021,dalacu_nanowire-based_2019} are appreciated for the absence of PL lines from neighboring emitters, which must be scrutinized with care in the case of Stranski–Krastanov QDs. In addition, the presence of different exciton lines of the same QD may make the extraction of temperature more delicate, for instance in large II-VI or III-V QDs, as both the exciton binding energies and the phonon sideband intensity decrease when the QD size increases.

We nevertheless used the same procedure to analyze the phonon sidebands of other systems and materials, such as III-V quantum dots (GaAs \cite{peter_phonon_2004}, InAs \cite{favero_acoustic_2003,reigue_boite_2017}, InAsP \cite{dalacu_nanowire-based_2019}, GaN \cite{rol_probing_2007}), 2D emitters  (WSe2 \cite{vannucci_single-photon_2024}) and color centers (hBN \cite{khatri_phonon_2019}). We could measure the phonon temperatures of all these systems from the published PL spectra and the temperature obtained was consistently higher than the cryostat temperature used in the experiment. Two examples are described in the Supplementary Information: they illustrate the dramatic effect of changing the excitation from non-resonant to quasi-resonant. Note that Stranski–Krastanov QDs embedded in a broad matrix are expected to be slightly less sensitive to heating, but nevertheless suffer from the low thermal conductivity at cryogenic temperatures. Also, any structuration around the QDs, as well as implantation defects, are likely to decrease the phonon mean free path and thus to decrease the thermal conductivity. 

Heating the QD shifts the ZPL position, as shown in Fig.\ref{fig:fig2}. The thermal redshift at constant excitation power is shown in Fig.~\ref{fig:energy-shift} of Section Methods, and modelled using the description of the bandgap evolution by Passler \cite{passler_parameter_1999}. The redshift observed upon increasing the excitation power at constant cryostat temperature closely follows the thermal shift, provided the two are plotted as a function of the phonon temperature. Indeed, this approach can be used to estimate the local temperature,\cite{misiara_effects_2024,dharmasiri_sensitivity_2024,gao_local_2024} but it requires calibration. The same applies to an estimate of the temperature from the temperature dependence of the Huang and Rhys factor (ZPL relative intensity). In the present study, the three methods, phonon sideband ratio, ZPL shift, and (not shown) ZPL relative intensity, conclude to the same value of the laser heating. 

\section{Conclusion}

We have introduced a new method to measure the local temperature of acoustic phonons coupled to a single photon emitter, a key parameter for its use in quantum information. Analyzing the ratio of Stokes and anti-Stokes sidebands allows for directly measuring the absolute phonon temperature without requiring external calibration. Importantly, this simple method is broadly applicable, as it is independent of the specific coupling mechanisms or the phonon density of states.

This technique was demonstrated with a single CdSe QD embedded in a ZnSe nanowire, allowing us to measure a uniform temperature of all acoustic phonons coupled to the emitter. Measurements were conducted across a wide temperature range, from $6\,\mathrm{K}$ to $100\,\mathrm{K}$. Additionally, significant heating of the QD was observed under non-resonant excitation, attributed to the material's low thermal conductivity at cryogenic temperatures and the sample geometry.

The method's versatility has also been demonstrated on other systems, including III-V quantum dots, two-dimensional emitters, and color centers, highlighting the importance of employing optimized system design and excitation strategies to mitigate heating effects. This technique thus provides a simple tool to probe the local temperature and optimize emerging single emitters. It also opens up new avenues for developing primary thermometers based on quantum emitters.

\section{Methods}
\label{sec:methods}

\subsection{Experimental setup}
\begin{figure}
  \centering
   \includegraphics[width=0.5\textwidth]{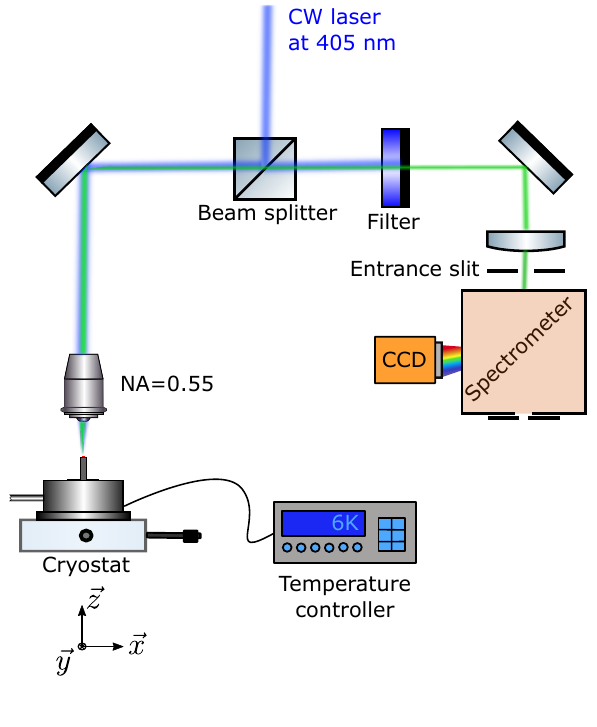}
    \caption{
Schematic of the spectroscopy setup used for the temperature-dependent measurements.}
  \label{fig:pl_setup}
\end{figure}

The sample is placed under vacuum and positioned on top of a continuous-flow cold-finger cryostat, capable of rapidly cooling the sample to 5-6 K for cryogenic PL measurements. To maintain precise temperature control, the temperature of the cryostat finger is constantly monitored using a dedicated temperature controller module. The patterned substrate and the X-Y linear stage equipped with piezoelectric movements allow us to precisely track the emitter’s position while adjusting the cryostat temperature. A schematic representation of the spectroscopy setup is provided in Fig~\ref{fig:pl_setup}. 

Within this experiment, we employed a continuous wave laser operating at a wavelength of 405 nm, corresponding to a photon energy of 3.06 eV. This is larger than the bandgap of ZnSe, which is 2.70 eV at 300K and 2.82 eV at 10K.  The QD-NWs are excited \textit{via} a microscope objective with a numerical aperture of 0.55. The single photons emitted by the QD are guided by the nanowire shell and collected along the same axis in the microscope. The spectrometer (0.46 m focal length) is equipped with a high-resolution 1800 grooves/mm grating and an entrance slit width of 0.05 mm. This configuration provided a spectral resolution of better than 200~µeV, for precise spectral line measurements on the CCD camera. 

\subsection{Temperature extraction}

Careful processing is essential to extract the phonon bath temperature from raw PL spectra. The reliability of the temperature extracted depends significantly on this process.

First, a constant baseline is subtracted from the spectra. This step is straightforward when the spectrum features only a single line, but it becomes more intricate when additional lines or emitters are present. Note that the wavelength range recorded must encompass the whole phonon sideband.

\begin{figure*}
  \centering
  \includegraphics[width=\textwidth]{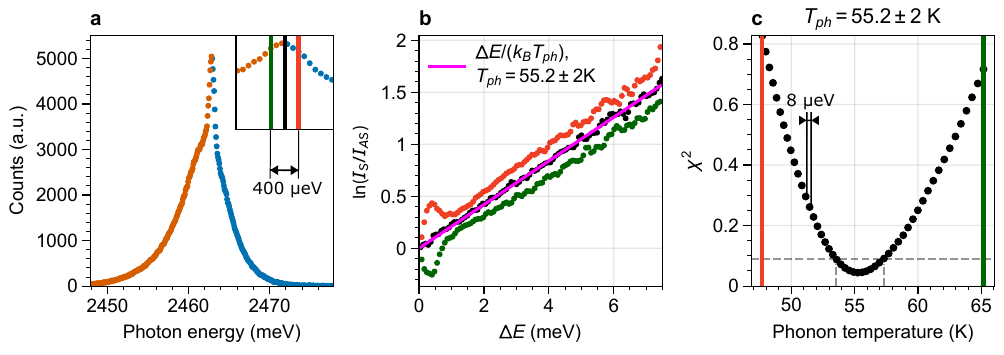}
  \caption{(a) Intensity spectrum at $T_\mathrm{cryo}=50\,\mathrm{K}$ and $P=14\,\mu\mathrm{W}$. Stokes ($I_S$, orange) and anti-Stokes ($I_{AS}$, blue) intensities are separated by the expected ZPL position (black line). Green and red lines mark $\pm200\,\mu\mathrm{eV}$ from the ZPL. (b) Sidebands intensity ratio $I_S/I_{AS}$ fitted with $\Delta E/(k_\mathrm{B}T_{\mathrm{ph}})$ (magenta line shows the best fit). (c) $\chi^2$ vs. phonon temperature.}
  \label{fig:method_extraction}
\end{figure*}

Then, the central position of the ZPL must be precisely determined. This position deeply affects the computation of the Stokes and anti-Stokes sidebands and, thus, the extracted temperature. In Fig.~\ref{fig:method_extraction}, adjustments of the ZPL position by a few $\mu\mathrm{eV}$ ($8~\mu$eV in that case) are made to find the accurate temperature. If the ZPL is offset from its position, the plot of $\ln(I_S/I_{AS})$ deviates from the linear evolution around the ZPL, as observed in Fig.~\ref{fig:method_extraction}b.

The central energy of the ZPL likely does not coincide with any data points. An interpolation of the Stokes and anti-Stokes sidebands is necessary to ensure that the energies of each sideband align, allowing the ratio of the two to be computed at the same energy shift, $\Delta E$.

Finally, for each ZPL position, a phonon temperature is extracted from the fit of the data using the expression $\Delta E/(k_\mathrm{B}T_{\mathrm{ph}})$ where the quality of the fit is given by $\chi^2$.

\begin{equation}
    \chi^2=\sum_i^N r_i^2,
\end{equation}

with $r_i$ the residual values returned from the fit and N the number of data points.

The uncertainty in the extracted temperature is determined at $2\chi^2_{\text{min}}$. In Fig.~\ref{fig:method_extraction}c, $\chi^2_{\text{min}}=0.04$ is obtained at $E_{ZPL}=2462.878$ meV (black vertical line in Fig.~\ref{fig:method_extraction}a) for a phonon temperature of $T_{ph}=55.2\pm2$ K.

\subsection{Radiation losses}

The radiation losses $P_r$ can be estimated using the Stefan-Boltzmann law $P_r = \epsilon A_S \sigma \left( T_{\text{ph}}^4 - T_{\text{cryo}}^4 \right)$, where $A_S$ is the surface area of the NW, $ \epsilon $ is the emissivity (always less than 1), $T_{\text{ph}}$ the temperature at the level of the QD, and $ \sigma $ is the Stefan–Boltzmann constant. For a cylindrical NW of length $10 \, \mu\text{m} $ and diameter $ 200 \, \text{nm} $, with $ T_{\text{ph}} = 300 \, \text{K} $ and $ T_\mathrm{cryo} = 6 \, \text{K}$, the maximum radiation loss power is estimated as $P_{r, \text{max}} \approx 0.003 \, \mu\text{W}$. This is negligible compared to the power absorbed by the NW.

\subsection{Derivation of the method}

The interaction of an electron or exciton with phonons can be exactly solved within the Independent Boson Model \cite{mahan_electronphonon_2000}. 

Without loss of generality, we consider a QD interacting with a set of $N$ discrete acoustic-phonon modes, denoted as $q_i$, each with energy $\hbar\omega_i$ and population $n_i$ at thermal equilibrium. The population $n_i$ follows the Bose-Einstein distribution:

\begin{equation}
n_i = n(\hbar\omega_i, T) = \frac{1}{e^{\hbar\omega_i/k_BT} - 1}.
\end{equation}

Using the ZPL energy $E_{ZPL}$ as the reference energy, the spectral function is expressed as \cite{besombes_acoustic_2001}:

\begin{equation}
I(E) = \sum_{p_1, \ldots, p_N} W_{p_1} \ldots W_{p_N} \delta(E + \Delta E_{p_1, \ldots, p_N}),
\end{equation}

where the energy shift $\Delta E_{p_1, \ldots, p_N}$ is given by:

\begin{equation}
\Delta E_{p_1, \ldots, p_N} = \sum_{i=1}^N p_i \hbar\omega_i.
\end{equation}

The probability $W_{p_i}$ that the optical transition involves $p_i$ phonons in mode $q_i$ is:

\begin{equation}
W_{p_i} = \left(\frac{n_i + 1}{n_i}\right)^{\frac{p_i}{2}} e^{-g_i(2n_i + 1)} I_{p_i}(2g_i\sqrt{n_i(n_i + 1)}),
\end{equation}

where $I_{p_i}$ is the modified Bessel function of the first kind, satisfying $I_{-p_i} = I_{p_i}$, and $g_i$ is the coupling constant between the exciton and the phonon mode $q_i$.

Positive $p_i$ values correspond to emission of phonons while negative $p_i$ correspond to absorption of phonons. A multi-phonon process $\{p_1, \ldots, p_N\}$ giving $\Delta E_{p_1, \ldots, p_N} > 0$ leads to an optical emission in the Stokes sideband (at energy $-\Delta E_{p_1, \ldots, p_N}$ from the ZPL). 

Now, we consider the symmetrical multi-phonon process with $\{-p_1, \ldots, -p_N\}$, it leads to an optical emission in the anti-Stokes sideband (at energy $ +\Delta E_{p_1, \ldots, p_N}$). The ratio of the probabilities $W_S = \prod_i^N W_{p_i}$ and $W_{AS} = \prod_i^N W_{-p_i}$ of these two processes is:

\begin{multline}
\frac{W_S}{W_{AS}} = \prod_i^N \frac{W_{p_i}}{W_{-p_i}} = \prod_i^N \left(\frac{n_i + 1}{n_i}\right)^{p_i} \\
= \exp\bigg(\frac{\Delta E_{p_1, \ldots, p_N}}{k_BT}\bigg)
\end{multline}

The PL intensity of a Stokes process at $\Delta E$ result from all multi-phonon processes $j$ leading to the shifted energy $\Delta E$: 

\begin{align}
    I_S(\Delta E) &= \sum_j W_{S,j}(\Delta E) \nonumber \\
                  &= \sum_j e^{\Delta E/k_BT} W_{S,j}(\Delta E) \nonumber \\
                  &= e^{\Delta E/k_BT} I_{AS}(\Delta E)
\end{align}

Thus for $\Delta E > 0$,

\begin{equation}
\frac{I_S(\Delta E)}{I_{AS}(\Delta E)} = e^{\Delta E/k_BT} = \frac{I(\Delta E)}{I(-\Delta E)}.
\end{equation}

\subsection{Energy shift}
\label{sec:energy-shift}

\begin{figure}
  \centering
   \includegraphics{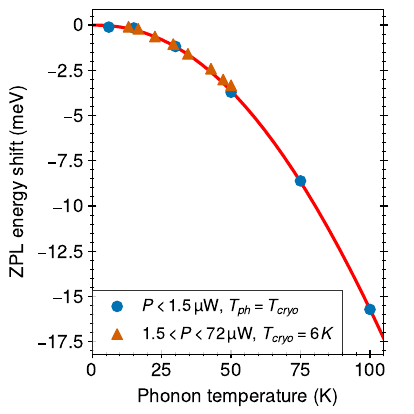}
    \caption{Energy shift of the ZPL as a function of phonon temperature for a CdSe/ZnSe QD-NW. Blue dots represent measurements at low excitation power ($P < 1.5\,\mu$W) for various cryostat temperatures $T_{\mathit{cryo}}$. Orange triangles correspond to measurements taken at different excitation powers ($1.5 < P < 72\,\mu$W) while maintaining a constant cryostat temperature of $T_{\mathit{cryo}} = 6$\,K. The red solid line represents a fit using the Pässler model.}
  \label{fig:energy-shift}
\end{figure}

The temperature dependence of the energy of the ZPL is presented in Fig.~\ref{fig:energy-shift} under two conditions. We first report the ZPL position under low excitation power excitation ($P<1.5$ $\mu\text{W}$) for various cryostat temperatures (blue dots). The horizontal scale is the phonon temperature $T_{ph}$ determined through the Stokes-to-antiStokes ratio. Then, the ZPL energies obtained under different excitation powers and $T_{cryo}=6$K from Fig.\ref{fig:fig3}c are represented with orange triangles.\\

The ZPL shows a redshift of about 15 meV as the temperature increases from 0K to 100K. These energy shifts ($E_{ZPL}(T_{ph})-E_{ZPL}(0K)$) are fitted using the band gap variation introduced by Pässler \cite{passler_parameter_1999}:

\begin{equation}
    E_{ZPL}(T_{ph}) = E_{ZPL}(0)-\frac{\alpha\Theta_p}{2}\bigg[\sqrt[p]{1+\bigg(\frac{2T_{ph}}{\Theta_p}\bigg)^p}-1\bigg].
\end{equation}

In this expression, the parameter $\alpha$ (meV/K) represents the $T_{ph} \rightarrow \infty$ limit of the slope. The quantity $\Theta_p$ (K) is an effective phonon temperature, and the fractional exponent $p$ is related to the phonon dispersion of the material. From the fit of data with $P<1.5\mu$W shown in Fig.~\ref{fig:energy-shift}, we obtained $E_{ZPL}=2467.2$ meV, $\alpha=0.65$ meV/K, $\Theta_p=313$ K and $p=2.206$.

The data obtained under the two conditions feature the same dependence, confirming that this method can be used as an alternative for measuring the local temperature of a single emitter. However, unlike the method described in the text, this approach requires calibration. In addition, its accuracy is limited at low temperature.

\begin{acknowledgement}

The authors thank O. Bourgeois, J. Claudon and J-M Gérard for fruitful discussions. We also thank V. Voliotis and J. Ferreira Neto for providing and explaining their data.\\

We acknowledge funding from LANEF in Grenoble, ANR-10-LABX-51-01, and CEA-PE Bottom-Up QPhotonics.

\end{acknowledgement}

\begin{suppinfo}

        Application of the method to two other systems: $\text{WSe}_2$ quantum emitter \cite{vannucci_single-photon_2024} and InAs/GaAs quantum dot \cite{reigue_boite_2017}.

\end{suppinfo}

\bibliography{references}

\end{document}